\documentclass[english,a4paper,12pt]{article}
\usepackage[T2A]{fontenc}
\usepackage[cp1251]{inputenc}
\usepackage{babel}
\usepackage{amsmath}
\usepackage{graphicx}
\usepackage{amssymb}
\usepackage{epsf}
\usepackage{pazhe}
\usepackage[labelsep=period,hang]{caption} 

\tightenlines

\newcommand{\WI}[2]{#1_{\mathrm{#2}}}
\newcommand{\rhob}{\WI{\rho}{b}}

\newcommand{\mb}{\WI{m}{b}}

\sloppypar

\begin{document}
	\baselineskip 21pt
	
	
	\title{\bf Thermal Neutrinos from the Explosion of a Minimum-Mass Neutron Star}
	
	\author{\bf \hspace{-1.3cm} \
		A.V. Yudin\affilmark{1*}, N.V. Dunina-Barkovskaya\affilmark{1}, S.I. Blinnikov\affilmark{1}}
	
	\affil{
		{\it National Research Center ``Kurchatov Institute'', Moscow, 123182 Russia}$^1$}
	
	\vspace{2mm}
	
	\sloppypar
	\vspace{2mm}
	\noindent
	We present our calculations of the thermal neutrino radiation that accompanies the explosion of
a minimum-mass neutron star. In this case, the neutrino luminosity is lower than the luminosity during a
supernova explosion approximately by five orders of magnitude, while the energy carried away by neutrinos
is low compared to the explosion energy. We also show that the energy losses through neutrinos do not
hinder the envelope heating and the cumulation of the shock during its breakout and the acceleration of the
outer part of the envelope to ultrarelativistic speeds.
	

	\vfill
	\noindent\rule{8cm}{1pt}\\
	{$^*$ email: $<$yudin@itep.ru$>$}
	
	\clearpage
	
\section*{INTRODUCTION}
\noindent In 2017 the simultaneous detection of the
gravitational-wave signal GW170817 and the gamma-ray
burst GRB170817A (Abbott et al. 2017) confirmed
the long predicted connection (Blinnikov
et al. 1984) between short gamma-ray bursts and the
merging of neutron stars (NSs). The coalescence
process itself usually appears as the merging of two
NSs into one object, a supermassive NS or a black
hole, with a smallmass ejection (for a review, see, e.g.,
Baiotti and Rezzolla (2017) and references therein).
However, there exists a competing mechanism of the
process, stripping, within which the more massive
NS devours its light companion (Clark and Eardley
1977). Referring the interested reader to recent
papers on this subject (Blinnikov et al. 2021, 2022),
we will dwell here only on the most important component
of this mechanism, the explosion of a minimummass
neutron star (MMNS). The explosion process
itself, first calculated by D.K. Nadyozhin (Blinnikov
et al. 1990), turned out to be close in energetics to
the classical supernova explosion energy, $10^{51}$~erg.
In this case, the outer stellar layers are heated to
temperatures $T \simeq 10^9-10^{10}$~K, producing a weak
gamma-ray burst with an energy $10^{43}{-}10^{47}$~erg,
in remarkable agreement with the observations of
GRB170817A.

The minimum-mass ($\sim 0.1M_\odot$) neutron star has
a peculiar structure: a tiny, $\sim 10$~km core containing
the bulk of the mass and an extended envelope
with a radius  $200{-}300$~km. Therefore, the general
relativity effects for it are relatively small. However,
during the MMNS explosion its matter acquires an
average speed 10\% 
of the speed of light $c$, while the
outer layers are accelerated to ultrarelativistic speeds.
In our recent paper (Yudin 2022) this process was
studied in terms of relativistic hydrodynamics and
the main properties of the explosion were shown to
remain approximately the same as those in the case
of using the nonrelativistic approach. At the same
time, the following question arose: Can the energy
losses through neutrino radiation from the hot outer
layers of an exploding NS reduce significantly their
temperature and/or hinder the envelope acceleration
to speeds of the order of $c$? Our paper gives an answer
to this question.

The paper is organized as follows. First we will
briefly describe the equations used to calculate the
MMNS explosion. Then we will discuss the thermal
neutrino radiation processes that we took into account.
Next we will present the main results obtained
in our simulations. In conclusion, we will compare
our results with the previously known ones and will
discuss what else should be done within the MMNS
explosion scenario.

\section*{BASIC EQUATIONS}
\noindent Let us give the basic equations of the problem
following Hwang and Noh (2016) and Yudin (2022).
We will place emphasis on those changes that are
caused by the inclusion of neutrino radiation. The
equations are written in Lagrangian coordinates for
the case where the problem is spherically symmetric.

The continuity equation is
\begin{equation}
\frac{1}{\rhob\gamma}=\frac{4\pi}{3}\frac{\partial r^3}{\partial\mb}.\label{Continuty}
\end{equation}
Here, $\rhob$ is the baryonic matter density, $r$ is the radius
(Eulerian coordinate), $\mb$ is the baryonic mass
(Lagrangian coordinate), and $\gamma$ is the Lorentz factor:
\begin{equation}
\gamma\equiv\frac{1}{\sqrt{1{-}v^2/c^2}},\label{gamma}
\end{equation}
where $v\equiv dr/dt$ is the matter velocity, and $t$ is the
coordinate (Schwarzschild) time.

The energy equation is written as
\begin{equation}
\frac{dE}{dt}+P\frac{d}{dt}\!\left(\!\frac{1}{\rhob}\!\right)=\frac{Q}{\rhob}\frac{d}{dt}\Big(\!\ln(\rhob r^3)\!\Big)-\frac{R_\nu}{\gamma}.\label{Energy}
\end{equation}
where $E$  is the internal energy of the matter per
unit mass, $P$ is the matter pressure, and $Q\geq 0$ is
the artificial viscosity the expression for which is
taken from Liebendoerfer et al. (2001) (for details, see
Yudin 2022). The last term in (\ref{Energy}) with $R_\nu\geq 0$ describes
the energy losses through neutrino radiation,
the factor $1/\gamma$ arises here due to the relativistic time
dilation.

The equation of motion (Euler equation) is
\begin{multline}
\frac{d}{dt}\Big(\gamma v\big[1+\frac{E{+}(P{+}Q)/\rhob}{c^2}\big]\Big)=\gamma^3\Big(1+\frac{E{+}P/\rhob}{c^2}\Big)\WI{a}{G}-\\
-4\pi r^2\frac{\partial(P{+}Q)}{\partial\mb}
-\frac{3Q}{\gamma r\rhob}-\frac{v}{c^2}R_\nu.\label{motion}
\end{multline}
Here, the first term on the right-hand side describes
the influence of the gravitational acceleration $\WI{a}{G}$ (see
below), the second term is attributable to the pressure
gradient, the third term is the additional contribution
from the artificial viscosity, and, finally, the last termis
determined by the influence of the neutrino radiation.

The Poisson equation for the gravitational potential
$\varphi$ can be rewritten in the form of an equation for
the acceleration $\WI{a}{G}=-\partial\varphi/\partial r$ as
\begin{equation}
\gamma\frac{\partial\left(r^2\WI{a}{G}\right)}{\partial\mb}=
{-}G\Big[2\gamma^2{-}1+\frac{1}{c^2}\Big((2\gamma^2{-}1)E+\frac{1}{\rhob}\big((2\gamma^2{+}1)P +2(\gamma^2{-}1)Q\big)\Big)\Big].\label{aG}
\end{equation}
The quantities $\WI{\rho}{b}, E, P, Q$ and $R_\nu$ refer to the comoving
frame, while $r$, $v$ and $\gamma$ refer to the laboratory one.

To close the system of equations (\ref{Continuty}-\ref{aG}), we need
to specify the energy losses through neutrino radiation
$R_\nu$, which will be done in the next section, and
the equation of state used. Regarding the latter (for
more details, see Yudin 2022) we will say that the
thermodynamic quantities (pressure, internal energy,
etc.) in this equation are the sum of several terms:
the main, temperature-independent contribution including
the nonideality of the matter (Haensel and
Potekhin 2004) and the temperature contribution of
the ideal gas and blackbody radiation.

Thus, all of the equations of the problem are completely
defined. The computations were performed on
a spherically symmetric Lagrangian grid with $\sim 3000$
cells. The equations written according to an implicit
scheme were solved by the matrix sweep method.

\section*{NEUTRINO ENERGY LOSSES}
\noindent In this section we will discuss the most significant
thermal energy losses of the matter through neutrino
radiation. In this case, we will follow mainly
Itoh et al. (1996). The subroutines in \textsf{FORTRAN}
computing the various contributions to the total
energy loss rate $R_\nu$ discussed below are accessible
at \verb"https://cococubed.com/code_pages/nuloss.shtml".
We will use them in our computations.

	\begin{figure}[htb]
		\begin{center}
			\includegraphics[width=15cm]{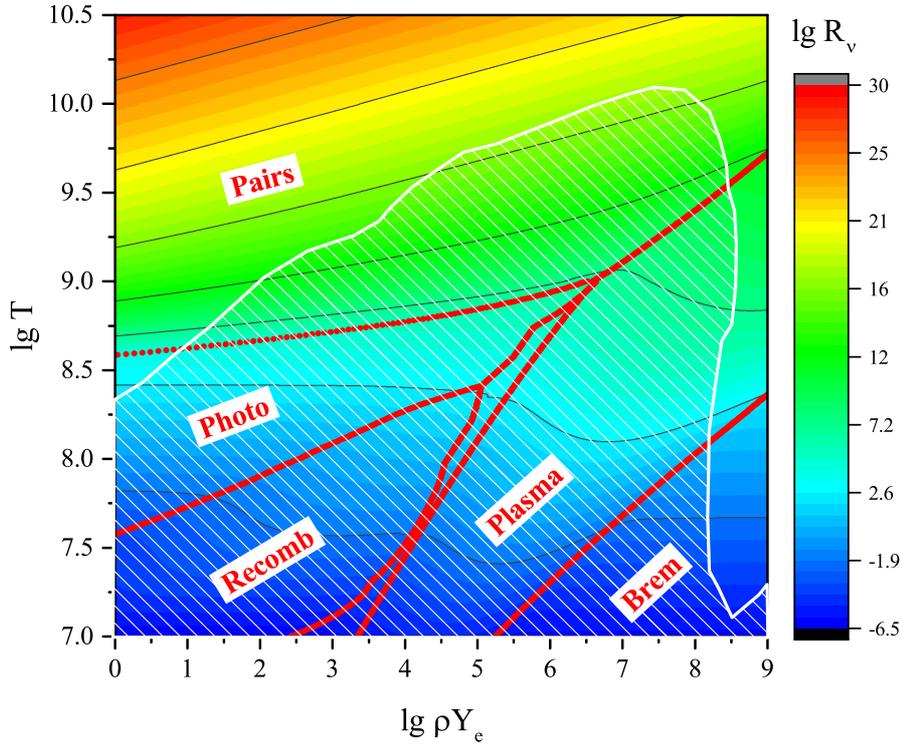}
		\end{center}
\vspace{-1cm}
		\caption{Isolines of the energy losses through neutrinos $R_\nu$($\mbox{erg}/\mbox{g}/\mbox{s}$) (right color panel) on the $\lg T[\mbox{K}] -\lg\WI{Y}{e}\rho[\mbox{g}/\mbox{cm}^{3}]$
plane. The white hatching indicates the zone of parameters encountered in our previous NS explosion computations. The red
lines highlight the regions of dominance of a particular process --- Pairs: pair annihilation, Photo: photo-neutrinos, Plasma:
plasmon decay, Recomb: neutrinos from recombination, Brem: bremsstrahlung.} \label{pix_rho_T_nu}
	\end{figure}
Isolines of $R_\nu$ for the matter composition corresponding
to $^{56}\mathrm{Fe}$ are shown in Fig.~\ref{pix_rho_T_nu} on the
$\lg T{-}\lg\rho\WI{Y}{e}$ ($\WI{Y}{e}$ is the electron-to-baryon ratio
in the matter) plane. The red lines indicate the
boundaries of the regions corresponding to the dominance
of one particular radiation mechanism (the
regions are labeled, the narrow band between the
recombination and plasma neutrinos corresponds to
bremsstrahlung). The white hatching indicates the
zone of parameters encountered in our previous computations
of the explosive disruption of a minimum mass
NS (Yudin 2022). As can be seen, the losses
through pair radiation as well as the plasmon decay
and the photo-neutrino reaction dominate in the
region of maximum neutrino losses; the remaining
processes are important at lower temperatures. Let
us consider these processes in more detail.

The neutrino losses caused by the electron–
positron pair annihilation $e^{-}+e^{+}\rightarrow\nu+\tilde{\nu}$ (annihilation
losses) dominate in the part of the $\lg T -\lg\rho\WI{Y}{e}$
plane where enough positrons are produced due to a
high temperature (of the order of the electron mass, in
units of $k=c=1$) and low degeneracy. These losses
were calculated by Beaudet et al. (1967), but without
taking into account the neutral currents responsible
for the production of neutrino flavors different from the
electron one. In Itoh et al. (1996) the neutral currents
were already taken into account.

The neutrino losses caused by the plasmon decay
$\WI{\gamma}{plasm}\rightarrow \nu+\tilde{\nu}$ were also considered by Beaudet
et al. (1967), and these authors proposed their
approximation to calculate these losses, accurate
enough in the region $8 < \lg T < 10$ and $1 < \lg\rho <14$ (here, $T$ is measured in $K$ and $\rho$ is
in $\mbox{g}/\mbox{cm}^{3}$). Subsequently, Blinnikov and Dunina-
Barkovskaya (1994) proposed their approximation for
this process, more accurate in the range of lower
temperatures corresponding to the cooling of hot
white dwarfs. These authors also obtained an upper
limit on the neutrino magnetic moment on which
the neutrino losses caused by the decay of plasmons
depend. An approximation with a wide range of
applicability and a high accuracy was also proposed
by Kantor and Gusakov (2007).

The neutrino losses caused by the photo-neutrino
reaction $e^{\pm}+\gamma\rightarrow e^{\pm}+\nu+\tilde{\nu}$ were considered by
Petrosian et al. (1967), but without taking into account
the plasma corrections that were subsequently
applied by Beaudet et al. (1967). Itoh et al. (1996)
also took into account the neutral currents.

The neutrino losses caused by the bremsstrahlung
 $e^{-}+(A,Z)\rightarrow e^{-}+(A,Z)+\nu+\tilde{\nu}$ depend not only
on the ratio $\WI{Y}{e}=Z/A$, but also directly on the nuclear
charge $Z$. For example, it can be seen in Fig.~6 from
Itoh et al. (1996) that at $\lg T = 7$ and $\lg\rho\WI{Y}{e}=12$
the neutrino losses through bremsstrahlung for $^{56}\mathrm{Fe}$
are higher than those for $^{12}\mathrm{C}$ by almost five orders of
magnitude (see also Ofengeim et al. 2014).

The neutrino losses caused by the recombination $\WI{e^{-}}{contin}\rightarrow\WI{e^{-}}{bound}+\nu+\tilde{\nu}$ also depend on the nuclear
charge $Z$ at constant $\WI{Y}{e}=Z/A$ and exceed the other
losses in the region of comparatively low densities and
temperatures where the neutrino losses are small.

When computing the explosion, we assumed the
nuclear composition of each Lagrangian zone to be
fixed and equal to the initial MMNS composition (the
MMNS structure is shown, for example, in Fig.~2
from Blinnikov et al. (2021)). Given the above strong
dependence of some processes on the specific composition
(i.e., on $Z$, $\WI{Y}{e}$, etc.), this simplification is
important, although it should not affect qualitatively
the main results.

\section*{SIMULATION RESULTS}
\noindent The simulation of explosive MMNS disruption is
described in detail in Yudin (2022). Here, we will
outline the main stages only briefly. After the loss of
stability, the stellar expansion begins from the surface
and encompasses the entire star in a fraction of a
second. In this case, propagating along the descending
density profile, the acoustic vibrations turn into
weak shocks heating the envelope with a radius of
hundreds of kilometers. By about 0.3~s after the onset
of expansion, a strong shock is generated in the outer
part of the star that heats up the envelope to temperatures
$10^{10}$~K and accelerates it to ultrarelativistic
speeds. At the same time, the central part of the star
containing the bulk of its mass remains cool. The
further stellar expansion occurs virtually in the regime
of free expansion with a mean speed $0.1c$. As our computations showed, the inclusion of thermal
neutrinos does not change this picture.

\begin{figure}[htb]
		\begin{center}
			\includegraphics[width=10cm]{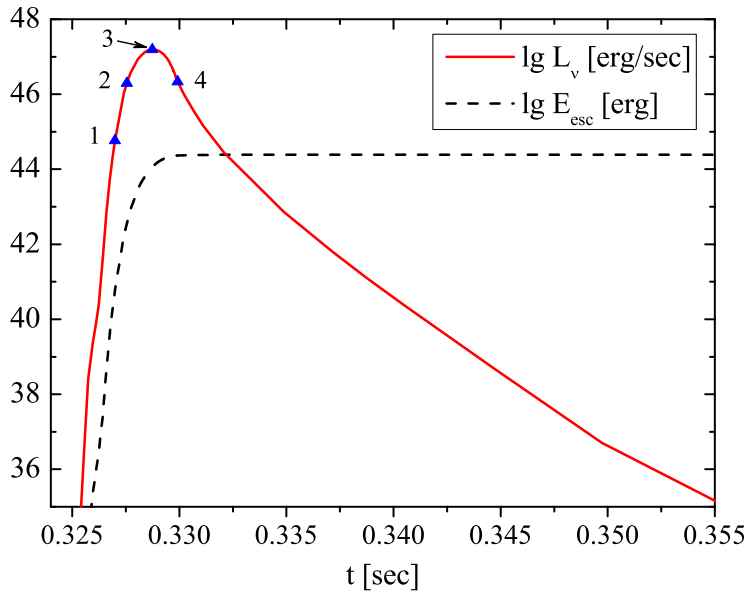}
		\end{center}
		\caption{The neutrino luminosity [$\mbox{erg}/\mbox{s}$] (red solid line) and the total energy [$\mbox{erg}$] carried away by neutrinos (black dashed line)
as functions of time. The blue triangles with numbers mark the four instants of time for which Fig.~\ref{pix_three_graphs} shows the distributions of
some parameters in the star.} \label{pix_L_and_E}
	\end{figure}
The main results concerning the properties of
the thermal neutrino radiation accompanying the
MMNS explosion are shown in Fig.~\ref{pix_L_and_E}. The red solid
line indicates the logarithm of the neutrino luminosity
$L_\nu$ [$\mbox{erg}/\mbox{s}$], while the black dashed line indicates
the logarithm of the total energy carried away by
neutrinos by a given time $\WI{E}{esc}$ [$\mbox{erg}$]. As can be
seen, the luminosity is much lower than that during
a supernova explosion ($10^{53}$~$\mbox{erg}/\mbox{s}$), while the
energy carried away is low compared to the kinetic
energy of the MMNS explosion ($10^{51}$ $\mbox{erg}$). This is
not surprising, since a very small part of the star is
subject to heating (see below).

\begin{figure}[htb]
		\begin{center}
			\includegraphics[width=\linewidth]{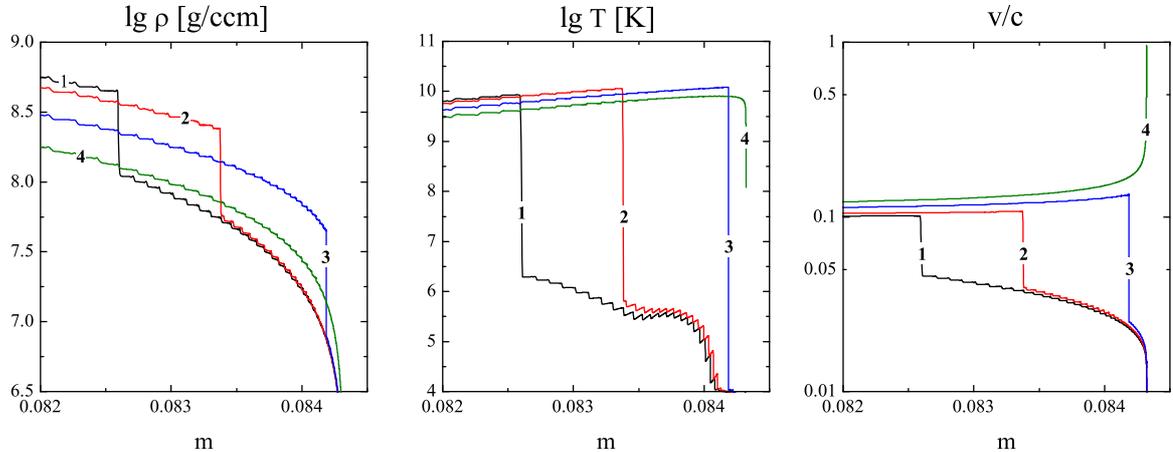}
		\end{center}
		\caption{From left to right: density $\rho[\mbox{g/cm}^3]$, temperature $T [\mbox{K}]$, and expansion velocity $v/c$ of the matter as functions of the
mass coordinate $m [M_\odot]$ for the four chosen instants of time shown in Fig.~\ref{pix_L_and_E}.} \label{pix_three_graphs}
	\end{figure}

To show the conditions accompanying the generation
of neutrino radiation, we chose four instants of
time near the luminosity peak; they are indicated by
the blue triangles with numbers in Fig.~\ref{pix_L_and_E}. The density,
temperature, and velocity distributions in the star at
these instants of time are shown in Fig.~\ref{pix_three_graphs} as functions
of the Lagrangian coordinate $m~[M_\odot]$ in the outer part
of the stellar envelope. The numbers on the curves
mark the instants of time. The neutrino luminosity
peak is seen to coincide with the shock breakout (instant
of time 3). On this graph we showed only those
regions of the star from which the bulk of the neutrino
radiation originates. As can be seen from Fig.~\ref{pix_rho_T_nu}, the
regions with a higher density are never heated enough
for the energy losses in them to be significant. In
contrast, the regions with a lower density (ahead of
the shock front) at these instants of time are still too
cool and, accordingly, the thermal neutrino radiation
there is also weak.

From the presented data it is easy to estimate
both the dominant type of the process responsible for
the neutrino radiation at this instant of time and the
fraction of the stellar mass from which this radiation
originates. Indeed, it can be seen from Fig.~\ref{pix_three_graphs} that near
the $L_\nu$ peak the temperature at the shock front is $T\sim 10^{10}$~K, while the density is $\rho\sim 10^{8}~\mbox{g}/\mbox{cm}^3$. It is clear
from Fig.~\ref{pix_rho_T_nu} that this is the region of dominance of the
electron–positron pair annihilation radiation, while
the power of the losses is $R_\nu\sim 10^{16}$ erg/g/s.
Hence the size of the region from which the neutrinos
originate is
\begin{equation}
\triangle m [M_\odot]=\frac{L_\nu}{R_\nu M_\odot}\sim 10^{-3},
\end{equation}
i.e., mainly the relatively narrow zone behind the
shock front radiates.

Another question concerns the applicability of the
neutrino radiation ``leakage'' scheme used by us that
allows it to be taken into account only in the bulk
energy losses, without considering the transfer processes.
The characteristic neutrino–matter interaction
cross section is (see, e.g., Burrows and Thompson
2004) $\sigma_\nu\sim 10^{-43}(E_\nu/\mbox{MeV})^2~\mbox{cm}^2$. For the dominant
type of the radiation process, pair annihilation,
the estimate (Misiaszek et al. (2006) for the neutrino
energy $E_\nu\sim 4 kT$ is valid in the ranges of densities
and temperatures under consideration. Hence, for
$\rho\sim 10^{8}~\mbox{g}/\mbox{cm}^3$ and $T\sim 10^{10}~\mbox{K}\sim 1~\mbox{MeV}$ the neutrino mean free path is
\begin{equation}
l_\nu\sim\frac{\WI{m}{u}}{\rho\sigma_\nu}\sim 10^{10}\mbox{cm},
\end{equation}
where $\WI{m}{u}$ is the atomic mass unit. Obviously, in
this case, treating the neutrinos as freely escaping
from the system, without taking into account the
probability of their recurrent interaction with matter,
is quite justified.

\section*{DISCUSSION AND CONCLUSIONS}
\noindent The main motivation for this study was to check
whether the thermal losses through neutrino radiation
could seriously affect the MMNS explosion and, in
particular, the cumulation of the shock during its
breakout. As our calculations showed, the effect from
these losses is small and does not lead to any significant
change in the explosion parameters. Nor does it
lead to any noticeable reduction in the temperature of
the stellar envelope (cf. Yudin 2022).

However, we everywhere emphasized with reason
that in our calculations we deal only with the \emph{thermal}
neutrinos emitted from the outermost regions of
the NS. In fact, during the disruption of a MMNS
its strongly neutronized matter experiences explosive
decompression whereby the nucleosynthesis processes
(the r-process; see Panov and Yudin 2020),
which are reduced, to a first approximation, to the
capture of neutrons by nuclei and their subsequent
beta decay, proceed. In this case, naturally, neutrinos,
predominantly $\WI{\tilde{\nu}}{e}$, are emitted and the radiation
comes from the entire mass of the matter rather than
from a narrow layer near the surface, as in the case
considered by us. Note in passing that here it is
important to take into account not only the energy
losses through neutrino radiation, but also the heating
caused by nuclear transformations in the matter.

Only a few calculations of such processes have
been carried out so far: in Colpi et al. (1989) and
Sumiyoshi et al. (1998). Of special note is the last paper,
in which hydrodynamic simulations of a MMNS
explosion with nonthermal neutrino losses were performed.
The peak luminosities obtained by the authors
are higher than those calculated by us approximately
by four orders of magnitude ($10^{52}~\mbox{erg}/\mbox{s}$),
while the explosion itself is extended in time and not
as powerful as that in our calculations ($10^{49}$ erg in
the kinetic energy of the matter rather than $10^{51}$ erg;
Yudin 2022). Unfortunately, as the authors themselves
admit, they used a moderately realistic equation
of state for the matter and an oversimplified description
of nucleosynthesis and the accompanying
neutrino losses. Therefore, it is now hard to say how
much their conclusions correspond to reality.

The essentially one-dimensional nature of
a MMNS explosion (Manukovskii 2010) gives hope
that it will be possible to avoid all of the numerical
difficulties associated with 3D NS merging simulations
when studying it. At the same time, within the
1D MMNS explosion model it is possible to study in
much more detail the phenomena occurring in this
case, in particular, to properly calculate the r-process
and the accompanying neutrino signal. At present
we are working on such calculations that take into
account the evolution of the nuclear composition of
matter and all of the possible, and not only thermal,
neutrino processes.

\vspace{1cm}
 {\bf ACKNOWLEDGMENTS}
We are grateful to the anonymous referees for their
valuable remarks.

The work of A.V. Yudin on the calculations of a
NS explosion was supported by RSF grant no. 21-12-00061. S.I. Blinnikov is grateful to RSF grant
no. 19-12-00229-P for supporting the development
of the approximations of neutrino radiation from
GRB170817A-type objects.

	\pagebreak

\end{document}